\documentclass[twocolumn,amsmath,amssymb,prl]{revtex4}

\usepackage{graphicx}
\usepackage{dcolumn}
\usepackage{bm}

\begin{document}



\title{Ideal glass transition in a simple 2D lattice model}

\author{Z. Rotman}
\author{E. Eisenberg}

\affiliation{Raymond and Beverly Sackler School of Physics and Astronomy,
Tel Aviv University, Tel Aviv 69978, Israel}



\begin{abstract}
We present a simple lattice model showing a glassy behavior. $R$
matrix analysis predicts critical termination of the super-cooled
fluid branch at density $\rho_g=0.1717$. This prediction is
confirmed by dynamical numerical simulations, showing power-law
divergences of relaxation time $\tau_{1/2}$, as well as the 4-susceptibility
$\chi_4$ peak's location and height exactly at the
predicted density. The power-law divergence of $\chi_4$
continues up to $\chi_4$ as high as $10^4$.
Finite-size scaling study reveals divergence of
correlation length accompanying the transition.
\end{abstract}


\maketitle

Understanding the transition of supercooled liquids into a glass is
considered by many to be one of the outstanding challenges of
condensed matter physics. Many liquids, when cooled fast enough to
avoid crystallization, appear to freeze into solid-like structures
devoid of crystalline order \cite{stillinger-n,benedetti,cavagna}.
The time scales for
structural relaxation in such metastable super-cooled regimes
increase dramatically as temperature is lowered.
For strong glass formers the relaxation times grow exponentially
$\tau=\tau_0\exp(A/T)$. Fragile glass formers exhibit relaxation times
that increase more rapidly than Arrhenius and are often fitted
by Vogel-Tammann-Fulcher (VTF) functional form
$\tau=\tau_0\exp[A/(T-T_0)]$, with a characteristic temperature
$T_0$ \cite{angell}.

The glass transition temperature is experimentally defined as the
temperature at which dynamic relaxation times exceed those
accessible in typical experiments, e.g. when viscosity hits
$10^{13}$ Poise. During the past century, much interest has been
focused on understanding the nature of this transition. Clearly, the
glass transition temperature, defined by some viscosity cutoff
value, is just an arbitrary reference point along the gradual
increase of relaxation times with decreasing temperature. The
question whether there is some deeper, profound, physical meaning to
the glass transition is still hotly debated \cite{stillinger-n}. Is
the fast increase of relaxation times merely a sharp cross-over in
the dynamics, or could it be manifestation of a true
thermodynamic transition? [Obviously, when considering a glass
transition in a system exhibiting a solid phase, such as real glass
formers, the notion of a thermodynamic glass transition must be
interpreted in the sense of a restricted part of phase space. For
simplicity, we ignore this distinction in the following.] Many
theoretical studies have been applied to support either one of the
competing views. For example, a popular microscopic approach is the
mode-coupling theory (see \cite{gotze,reichman} for reviews). It
predicts a dynamic glass transition, characterized by ergodicity
breaking, while thermodynamic (equilibrium) quantities such as the
isothermal compressibility do not become singular. In contrast, the
replica approach\cite{parisi} predicts a structural glass transition
with pure thermodynamic origin, characterized by a vanishing
configurational entropy. Other phenomenological theories, such as
the random first order transition (RFOT)\cite{kirk,wolynes} and the
potential energy landscape (PEL)\cite{sciortino} to name only two,
also predict a thermodynamic phase transition.

Recently, the $R$ matrix \cite{baramr,us} approach for analysis of the Mayer
cluster integrals expansion has been applied to the hard-spheres
fluid\cite{ebpre,eli}. It provides the density as a function of the activity
$z$ ($z=e^{\beta\mu}$, where $\mu$ is the chemical potential and $\beta$
is the inverse temperature) and predicts a critical termination of the
super-cooled fluid, with a power-law divergence of the isothermal
compressibility. The packing-fraction at which this
divergence is predicted to happen is 0.556(5), surprisingly close to
the experimentally reported glass transition packing-fraction 0.56(1).
This result, therefore, strongly supports the
existence of a thermodynamic glass transition for hard spheres
underlying the (experimentally and numerically observed) dynamical
arrest. It is desirable to have numerical measurements of the
super-cooled hard-spheres equation of state near the transition in
order to test the validity of the $R$ matrix approach. However,
these simulations are extremely challenging. Accordingly,
contradicting results have been reported regarding the existence of
singularities in thermodynamic quantities for this system
\cite{speedy, rintoul}.

The limits of numerical methods often hamper the study of glass
transition. Excluding the non-physical kinetically constrained
models, most models studied are either complex (binary
mixtures) or hard to simulate (hard spheres). They are therefore
limited in system size and simulation times. For example, a recent study of
Lennard-Jones binary mixture \cite{stein} reports that enlarging the
system to include $27000$ particles improves the quality of the
extrapolation of $k$-dependent quantities to zero wave vector.
Moreover, simulations are generally
limited to time scales roughly ten orders of magnitude shorter than
those near the laboratory glass transition temperature $T_g$ and
therefore to the initial stages of the glass formation process
\cite{berry}.
These numerical limitations might be lifted by introducing a simpler
model system that still captures the essence of glassy behavior.
Keeping that in mind we set to explore the glass transition is the
$N3$ lattice model.

The $N3$ model is a simple 2D model on a square lattice. Particles
interact only through hard-core exclusion up to the 3$^{\rm rd}$
nearest neighbor. The model is known to undergo a first order
solidification transition\cite{bellemans,orban,eliasher}, where
density jumps from $\rho_f\simeq0.161 $ to $\rho_s\simeq0.191$
\cite{eliasher} (the closest packing density is $0.2$).
Like the hard spheres case, $R$ matrix analysis predicts a critical
termination of the super-cooled fluid where the isothermal
compressibility power-law diverges. The critical density is found
to be $\rho_t\simeq0.1717$. In concordance with hard spheres results
\cite{eli}, we hypothesize that this point is indeed the
thermodynamic glass transition for this system. We then study the
dynamics of the model by extensive MC simulations and find that the
dynamical quantities diverge exactly at the density predicted. We
therefore conclude that the dynamical arrest in the $N3$ model
results from a singularity of the free energy, as predicted by the
$R$ matrix. These results support the view of a thermodynamic (a.k.a
ideal) glass transition in this system. Furthermore, we propose the
$N3$ system as a simple and convenient model-system for future
studies of glassiness.

\begin{table*}
\begin{tabular}{|c|c|c|c|}
\hline
$n$ & $nb_n$ & $B_n$ & $A_n$ \\
\hline
1 & 1                   & 13                     & 6         \\
2 & -13                 & 10.777777778                   & 5.4955088285   \\
3 & 205                 & 10.777970817                   & 5.4246225024   \\
4 & -3521              & 10.762751563                & 5.4025047989     \\
5 & 63466              & 10.755266974                & 5.3922495398     \\
6 & -1180075               & 10.751491280                 & 5.3866896951 \\
7 & 22423304               & 10.749147764                & 5.3834227892 \\
8 & -432957233            & 10.747459452              & 5.3814030748\\
9 & 8463267016            & 10.746108741             & 5.3801242739\\
10 & -167059758328            & 10.744940022           & 5.3793147595 \\
11 & 3323928207997            & 10.743879570          & 5.3788131550 \\
12 & -66571342665659         &             &  \\
13 & 1340690959181588         &            &  \\
14 & -27128411793067290      &        &  \\
15 & 551181809202093940      &         &  \\
16 & -11238651060745319617       &        &  \\
17 & 229877749269899350973       &          &  \\
18 & -4715081436294109369498    &  & \\
19 & 96953111901056596856377    &  & \\
20 & -1998044077291458477558756    &  & \\
21 & 41259643403438186795821307    &  & \\
22 & -853576114433438941428139775    &  & \\
23 & 17688270167244330924258385729    &  & \\

\hline
\end{tabular}
\caption{Mayer cluster coefficients $nb_n$ and R matrix diagonal
($B_n$) and off diagonal ($A_n$) elements for the N3 model}
\label{nbn}
\end{table*}

\begin{figure}
\includegraphics[width=6.9cm,height=6.9cm,angle=-90]{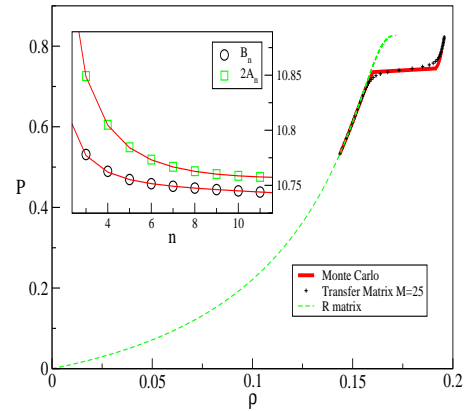}
\caption{(Color Online) N3 equation of state: $R$ matrix prediction,
based on the first $23$ Mayer cluster integrals (dashed line),
Monte-Carlo calculation on a $1000\times 1000$ lattice (solid line),
and exact transfer matrix calculation for a semi-infinite, $25$
sites wide, strip (symbols). The latter two methods provide
equilibrium results, while the $R$ matrix extrapolates to the
super-cooled fluid branch. The agreement of the $R$ matrix results
with the numerical methods is excellent throughout the fluid regime.
Inset shows the diagonal ($B_n$) and off-diagonal ($A_n$) $R$ matrix
elements, together with the fitted asymptotic form.} \label{fig-eos}
\end{figure}

In order to construct the $R$ matrix for the $N3$ model, we extended
the number of known Mayer cluster integrals to $23$, using the
transfer matrix (TM) method. We have employed a
diagonal-to-diagonal, symmetry reduced, TM, with strip width as
large as $M=24$ ($3874112$ symmetry-reduced classes). The cluster
integrals provide the exact $11\times11$ leading $R$ submatrix
presented in table \ref{nbn} (for details on $R$ matrix
construction, see \cite{us}). The matrix elements quickly converge
to a well-defined asymptotic form which we use to extrapolate
additional matrix elements and obtain the equation of state (figure
\ref{fig-eos}). Remarkably, the results, based only on low-density
expansion, are in an excellent agreement with both MC data and exact
TM calculations. The physical singularity is found at $z_t\sim
66.67$, well above the first order transition ($z_c\simeq 39.496$),
with a critical density $\rho_t=0.1717$. Furthermore, the $R$ matrix
provides an exact formula for the critical exponent $\sigma'$
associated with the termination point of the fluid \cite{eli,us}:
near this singularity, the density is given by
\begin{equation}
\rho_t-\rho(z) \simeq (z_t-z)^{\sigma'},
\end{equation}
and the critical exponent is found to be $\sigma'=0.39(2)$.

\begin{figure}
\includegraphics[width=6.9cm,height=6.9cm,angle=-90]{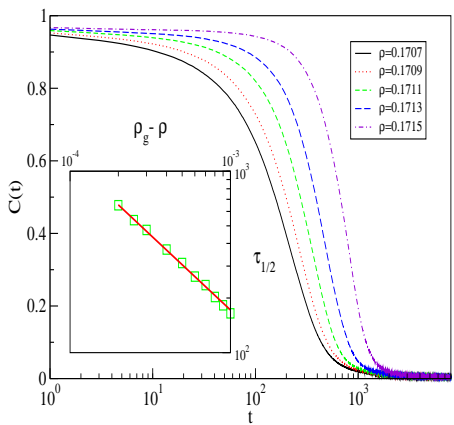}
\caption{(Color Online) Density-density auto-correlation \eqref{ct},
showing the typical glassy relaxation picture, on a logarithmic time
scale. The inset shows the relaxation time $\tau_{1/2}$ vs. density
(symbols), which is well-fitted by a power-law \eqref{tapl}
diverging at $\rho_g$.} \label{fig-tc}
\end{figure}

\begin{figure}
\includegraphics[width=6.9cm,height=6.9cm,angle=-90]{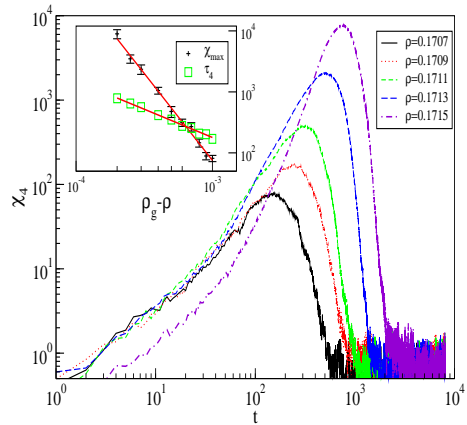}
\caption{(Color Online) $\chi_4$ as a function of time showing the
familiar peaks. Peaks' heights, $\chi_{\rm max}$, and locations,
$\tau_4$, power-law diverge as density approaches $\rho_g$ (inset)
with critical exponents 2.85 and 0.93 respectively.}
\label{fig-chi4}
\end{figure}

We hypothesize that this thermodynamic criticality underlies an
ideal glass transition for the N3 model, and set out to study the
model dynamically looking for signatures of this glass transition.
We conducted canonical (constant density) MC studies of the model in
the following way: The starting configuration was generated under
extreme cooling conditions (or, equivalently, infinite
chemical potential). Particles were allowed to diffuse when no
insertion was available. This process is known to terminate at the
random closest packing (RCP) state with density ~85\% of closest
packing density \cite{eli-rcp}. Here, we stop the cooling at the
desired density (below RCP), and let the system relax
diffusionally. Given enough time, the global equilibrium
phase-separation state is reached. On shorter time scale the system
relaxes to a disordered phase. We first measure the density-density
correlation
\begin{equation}
\label{ct} C(t)=\frac{1}{1-\rho}\left(
\frac{1}{N}\int\limits_{V}<n(r,0)n(r,t)>dr-\rho\right)
\end{equation}
along the relaxation process ($n(r,t)=1$ if a particle
exists at site $r$ at time $t$ and $0$ otherwise). Figure \ref{fig-tc}
shows the typical glassy dynamics picture: a plateau ($\beta$
regime) followed by a stretched-exponential decay ($\alpha$ regime).
Due to the discrete nature of the diffusion process in this model,
the $\beta$ relaxation stage is very short (of order
one simulation time unit) and is not presented. The relaxation time
$\tau_{1/2}$, defined as the time at which $C(t)=1/2$,
power-law diverges as the density approaches $\rho_g = 0.1717$:
\begin{equation}
\label{tapl} \tau_{1/2} (\rho) \sim (\rho_g - \rho)^{-\mu}
\end{equation}
with $\mu= 0.83$, (figure \ref{fig-tc}, inset). In addition, we
measure the 4-susceptibility
 $\chi_4$ \cite{dasgupta,toninelli}
\begin{equation}\chi_4(t)=N(<C(t)^2>-<C(t)>^2).\end{equation}
Again, a typical glassy behavior is observed (figure \ref{fig-chi4})
-- $\chi_4$ peaks at the $\alpha$ phase, and the peak grows in
height and shifts to higher times as density increases. Peak heights
($\chi_{max}$) and locations ($\tau_4$) also power-law diverge as
$\rho_g$ is approached (figure \ref{fig-chi4}, inset).
A transition to activated dynamics occuring close to the glass
transition could be manifested by the onset of a slower, logarithmic,
growth of the $\chi_4$ peak \cite{dalle}.
We do not observe any such transition for $\chi_4$ values up to $10^4$.

The above MC data confirm the $R$ matrix prediction to an
excellent agreement. Given that this prediction is
based solely on low-density series expansion, it is remarkable
that it captured quantitatively the behavior at the deep
super-cooled regime. This attests for the validity of the $R$ matrix
approach and its prediction of a thermodynamic
criticality in the equation of state of the N3 super cooled fluid, and
provides a strong evidence that the glass transition in this model
is indeed a thermodynamic, ideal, one.

\begin{figure}
\includegraphics[width=6.9cm,height=6.9cm,angle=-90]{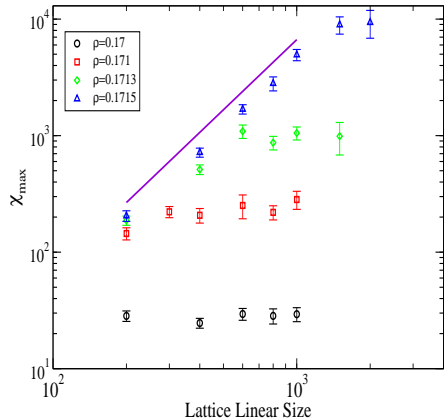}
\caption{(Color Online) Finite size analysis: $\chi_4$ peak height
as a function of lattice linear size, for various densities. As the
density approaches $\rho_g$, larger lattices are needed for
converged results, attesting for a diverging length scale. For
lattices smaller than the correlation length, $\chi_{\rm max}$ is
expected to grow like system's size. A straight line with a slope of
$2$ is presented, to guide the eye.} \label{fig-fss}
\end{figure}

The growing $\chi_4$ peak is indicative of growing cooperative
correlations in the relaxation process \cite{franz}. It measures the
volume upon which diffusional moves are
correlated\cite{doliwa}. In concordance, growing correlations
lengths are seen also by the emergence of finite size effects in the
density-density correlations as shown in figure \ref{fig-fss}. These
finite-size effects, recently highlighted by Karmakar et al.
\cite{karmakar} underscore the importance of using large systems for
MC studies of glassiness, which is most difficult for popular models
currently used.

Unlike the hard spheres case, the termination density $\rho_g$ predicted
by the $R$ matrix and the dynamical arrest occur in close proximity
to the random closest packing density $\rho_{rcp}$
\cite{eli-rcp}. Therefore, the region beyond the transition is inaccessible
in this model. It is important to note that the co-occurrence of the
two phenomena is not a universal trait of the $R$-matrix analysis.
For example, in the hard-spheres model the $R$ matrix prediction for
the ideal glass density is $\rho_t=0.556(5)$, much lower than
$\rho_{rcp}=0.64$. Hopefully, future work will find a model that
is as simple as the N3 but also allows access to densities beyond $\rho_t$.
This could be achieved by studying the soft-core N3 model, or other hard-core
lattice models.

We stress that the simplicity of the N3 model is important not only
in order to allow for analytical treatment, but to facilitate
numerical studies of large systems, much larger than those typically
used in glass studies. This is especially important when one
approaches the glass transition, where long-range cooperative
relaxation processes emerge, manifested by significant finite-size
dependence. For example, at density $\rho=0.1715$, even a
$1000\times 1000$ lattice ($171500$ particles; linear size $\sim
447$ particle diameters, much larger than typical 3D studies) is not
large enough to converge to bulk values as seen in figure
\ref{fig-fss}. The need for a simple model then is not a matter of
comfort but a real necessity. We therefore propose that the N3
model, or similar models, could serve in future studies of glass
formers being simple to handle, yet capturing the essence of
glassiness.

In conclusion, we have applied the $R$ matrix approach to the N3
model and found that its super-cooled equation of state becomes
singular at density $\rho_g=0.1717$, where the isothermal
compressibility power-law diverges. MC simulations
confirm that the model shows the characteristics of a fragile glass former
undergoing a glass transition at the predicted $\rho_g$.
It thus follows that in this model the phenomenological glass transition,
observed as a fragile-glass dynamical arrest at $\rho_g$,
is accompanied by a thermodynamic criticality.

\end{document}